\documentclass[twocolumn,tighten]{aastex61}
\pdfoutput=1 
\usepackage{amsmath,amstext}
\usepackage{apjfonts} 
\usepackage{afterpage}
\usepackage{xcolor}

\usepackage{graphicx}
\usepackage{epsfig}

\newcommand{\pixel}{\delta\theta}

\setwatermarkfontsize{100px} 

\DeclareFontFamily{U}{wncy}{}
\DeclareFontShape{U}{wncy}{m}{n}{<->wncyr10}{}
\DeclareSymbolFont{mcy}{U}{wncy}{m}{n}
\DeclareMathSymbol{\Sh}{\mathord}{mcy}{"58} 

 \shorttitle{Optimal Discretiization and Filtering}
\shortauthors{Psaltis et al}

\begin{document}


\title{Discretization and Filtering Effects on Black Hole Images Obtained with the Event Horizon Telescope}

\author{Dimitrios Psaltis}
\affiliation{Astronomy and Physics Departments, University of Arizona, 933 N. Cherry Ave, Tucson, AZ 85721, USA}

\author{Lia Medeiros}
\altaffiliation{NSF Astronomy and Astrophysics Postdoctoral Fellow}
\affiliation{School of Natural Sciences, Institute for Advanced Study, 1 Einstein Drive, Princeton, NJ 08540, USA}
\affiliation{Steward Observatory, University of Arizona, 933 N. Cherry Ave, Tucson, AZ 85721, USA}

\author{Tod R.\ Lauer}
\affiliation{NSF's National Optical Infrared Astronomy Research Laboratory, 950 North Cherry Ave., Tucson, AZ 85719, USA}

\author{Chi-kwan Chan}
\affiliation{Steward Observatory, University of Arizona, 933 N. Cherry Ave, Tucson, AZ 85721, USA}

\author{Feryal \"Ozel}
\affiliation{Astronomy and Physics Departments, University of Arizona, 933 N. Cherry Ave, Tucson, AZ 85721, USA}

\begin{abstract}
Interferometers, such as the Event Horizon Telescope (EHT), do not directly observe the images of sources but rather measure their Fourier components at discrete spatial frequencies up to a maximum value set by the longest baseline in the array.  Construction of images from the Fourier components or analysis of them with high-resolution models  requires careful treatment of fine source structure nominally beyond the array resolution.  The primary EHT targets, Sgr~A* and M87, are expected to have black-hole shadows with sharp edges and strongly filamentary emission from the surrounding plasma on scales much smaller than those probed by the currently largest baselines. We show that for aliasing not to affect images reconstructed with regularized maximum likelihood methods and model images that are directly compared to the data, the sampling of these images (i.e., their pixel spacing) needs to be significantly finer than the scale probed by the largest baseline in the array. Using GRMHD simulations of black-hole images, we estimate the maximum allowable pixel spacing to be $\simeq (1/8)GMc^{-2}$; for both of the primary EHT targets, this corresponds to an angular pixel size of $\lesssim 0.5~\mu$as. With aliasing under control, we then advocate use of the second-order Butterworth filter with a cut-off scale equal to the maximum array baseline as optimal for visualizing the reconstructed images. In contrast to the traditional Gaussian filters, this Butterworth filter retains most of the power at the scales probed by the array while suppressing the fine image details for which no data exist.
\end{abstract}


\section{INTRODUCTION}

The Event Horizon Telescope (EHT) is a Very-long Baseline Interferometric (VLBI) array operating at millimeter wavelengths, with baselines spanning the globe~\citep{PaperII}. Its main goal is imaging horizon-scale structures around supermassive black holes. The first images of the black hole in the center of M87 were announced in early 2019~\citep{PaperI, PaperIII, PaperIV, PaperV, PaperVI}.

As a sparse interferometric array, the EHT does not record images but rather various components of the complex visibilities at different baselines~\citep{PaperIII}. The largest possible baseline determines the effective resolution of the array. For the EHT observations of M87, the largest baseline achieved in 2017 between the IRAM 30m telescope in Spain and the JCMT/SMA station in Hawai'i was approximately equal to 8.2~$G\lambda$ at the observing wavelength of 1.3~mm, which corresponds to a resolution of $\sim 25~\mu$arcsec. The bright ring of emissions around the black hole shadow in M87 was measured to have a diameter of $42\pm 3~\mu$arcsec, which is  nominally only two resolution elements wide~\citep{PaperVI}. 

In order to generate black-hole images from the complex visibilities~\citep{PaperIV}, different image reconstruction algorithms were used based either on the traditional CLEAN method~\citep{Hogbom1974,Clark1980} or on regularized maximum likelihood methods (see, e.g.,~\citealt{Honma2014,Chael2016,Chael2018, Akiyama2017,Akiyama2018}). In a parallel approach, models for the complex visibilities that were based on geometric shapes~\citep{Kamruddin2013,Benkevitch2016} or General Relativistic Magneto-HydroDynamic (GRMHD) simulations~\citep{PaperV} were fit directly to the interferometric data to measure black-hole parameters and test the theory of General Relativity~\citep{PaperVI}.

Fundamentally, both the regularized maximum likelihood imaging methods and the GRMHD model comparisons are Bayesian parameter estimation methods. In both approaches, complex visibilities are calculated from models of discretized images via two-dimensional Fourier transforms, which are then compared to the data in order to obtain the most likely model parameters. In the case of the imaging methods, the model parameters are simply the values for the image brightness on each pixel on a two-dimensional grid. In the case of the GRMHD models, the model parameters are the physical properties of the black hole and the accretion flow (the mass, spin, orientation of the black hole, the electron density, temperature, etc); images are then generated by integrating the radiative transfer equation to calculate the image brightness on a similar two-dimensional grid of pixels.

In both approaches, the spacing of the pixels on the image defines a sampling frequency, which is often chosen arbitrarily. The Nyquist theorem states that structure with spatial frequencies half of the sampling frequency or less will be accurately represented, while more rapidly varying power will be aliased.  Aliasing occurs when high spatial frequencies that are present in the image beat against the sampling frequency and are incorrectly recorded as power at lower frequencies. Theoretical expectations on the sharpness of the black-hole shadow (see, e.g.,~\citealt{Psaltis2015}) and GRMHD simulations of the filamentary structure of the accretion-flow emission (see~\citealt{PaperV}) strongly suggest that the images are indeed dominated by prominent sharp structures at scales much smaller than their overall sizes. As a result, the Fourier transform of the images (i.e., the complex visibilities that the EHT actually measures) are expected to have substantial power at angular frequencies that are much larger than those probed by the largest baseline in the array. Consequently, evaluating the reconstructed and model images with pixel spacing that do not fully resolve these structures introduces aliasing errors to the calculated Fourier transforms that are then fit to the data.  Because the data for some of the EHT baselines have signal-to-noise ratios that exceed $\sim 100$~\citep{PaperIII}, comparing these biased Fourier transforms to the measured visibilities may further accentuate the biases in the inferred images or model parameters. Our first set of aims in this article is to use analytic considerations and results of GRMHD simulations to assess the biases introduced by the finite pixel spacing of the images and identify the pixel spacing that is optimal for the image structures expected around supermassive black holes.

Even though the above arguments (and the work reported below) suggest that the pixel spacing used in EHT images have to be much smaller than the nominal resolution of the array, the structures reconstructed at these small scales cannot be uniquely determined. In other words, many reconstructions that differ at small scales (large baselines) but agree with each other at large scales (small baselines) will be consistent with the data. For this reason, it is customary to report images only after they have been smoothed (convolved) by an elliptical Gaussian filter that suppresses Fourier frequencies larger than the largest baseline of the array (see, e.g., Figures~3 and 4 of \citealt{PaperI}). This convolution is, in fact, necessary for CLEAN methods, for which the model images are just a series of point sources.

Despite their ubiquitous use in interferometric imaging, Gaussian filters are suboptimal for this task, because they suppress Fourier power at all baselines, even the small ones for which data exist. In other words, images that have been smoothed by Gaussian filters in the traditional way are formally inconsistent with the interferometric data that they are meant to describe. The opposite extreme of a filter that would not suppress Fourier power at the observed baselines but sharply cuts off much larger scales, i.e., a circular top-hat filter, is not a solution to the problem either. The Fourier transform of such a filter has substantial ringing (sidelobes) that introduce bright spurious structures when convolved with the expected black-hole images characterized by sharp shadows. The second aim of this article is to devise a filter that does not suppress Fourier power at the observed baselines, that filters out the large baselines for which no data exist, and, at the same time, does not introduce spurious image artifacts.

\section{Definitions and General Considerations}

We define an image as a brightness distribution $I(x,y)$ in the sky, with $x$ and $y$ two angular coordinates typically oriented along the E-W and N-S orientations, respectively. We define a baseline vector $\vec{b}$ between two array sites as the vector that connects the two sites, projected orthogonally to the line of sight to a particular source. We also define the $(u,v)$ angular Fourier frequencies of the image along the E-W and N-S orientations respectively, i.e., we write the 2-dimensional Fourier transform of the image as
\begin{equation}
V(u,v)=\int\int e^{-2\pi(x u +y v)}I(x,y) dx dy\;.
\label{eq:vCZtheorem}
\end{equation}
The van Cittert-Zernike theorem states that $u$ and $v$ are equal to the two equivalent components of the baseline vector divided by the wavelength $\lambda$ of observation, i.e., that $(u,v)=\vec{b}/\lambda$. The Fourier transforms $V(u,v)$ are called the complex visibilities of the image. The EHT measures these complex visibilities at the baselines that are constructed by all possible pairs of stations in the array (see~\citealt{PaperIII}).
 
In image reconstruction or GRMHD model comparison algorithms, the model brightness in the sky is typically discretized on a two-dimensional grid of $N^2$ equidistant pixels, $I_{ij}$ with $i=1,...,N$ and $j=1,...N$. We denote by $\delta \theta$ the angular pixel spacing and by $\Delta\theta\equiv N\;\delta\theta$ the angular size of the entire image, i.e., the field-of-view.

It is often mathematically convenient  to write the discrete model of the image $I_{ij}$ as a function of a continuous set of variables in the form 
\begin{equation}
I_{\rm m}(x,y)=W_{\rm fov}(x;\Delta\theta) W_{\rm fov}(y,\Delta\theta) \Sh(x;\pixel) \Sh(y;\pixel) I(x,y) 
  \label{eq:discr}
 \end{equation}
(see~\citealt{TMS2017} as well as~\citealt{Pessah2007} for a similar approach in time-domain analysis). This continuous function becomes equal to the discrete model $I_{\rm m}(x_i,y_j)=I_{ij}$ on the discretization points, i.e., when $(x,y)=(x_i,y_j)$, and is equal to zero at all other places.

In equation~(\ref{eq:discr}), we have defined a number of useful functions. First is the pair of boxcar functions that determine the field of view along the $x-$ and $y-$axes; e.g., for the $x-$axis
\begin{equation}
W_{\rm fov}(x;\Delta\theta)\equiv \left\{\begin{array}{ll}
1, & {\rm if}\;,-\Delta\theta/2\le x\le \Delta\theta/2\\
0, & {\rm otherwise}
\end{array}\;,
\right.
\end{equation}
where the image is assumed to be zero-centered. Second is the pair of toothcombs of sampling functions along each of the axes; e.g., for the $x-$axis the Shah function 
\begin{equation}
\Sh(x;\pixel)\equiv\sum_{n=-\infty}^{\infty} \delta(x-n\; \pixel)
\end{equation}
samples the resulting image at the center of each pixel.

In words, equation~(\ref{eq:discr}) describes the discretization process as the product of the underlying continuous image with a two-dimensional toothcomb of $\delta$-functions (the grid of pixels) and a truncation by a two-dimensional boxcar function of width $\Delta\theta$ (the field-of-view). By the convolution theorem, the Fourier transform of the discrete image, which is also the map of complex visibilities, can be calculated as the convolution of the true visibility map of the image with the Fourier transform of the product of the field-of-view and sampling functions, $W_{\rm fov}$ and $S$, respectively. 

The Fourier transforms of the field-of-view functions are just sinc functions, i.e., for the x-direction
\begin{equation}
\tilde{W}_{\rm fov}(u;\Delta\theta)=\frac{\sin(\pi~u~\Delta\theta)}{\pi~u~\Delta\theta}\;,
\end{equation}
where we have used the tilde to denote Fourier transforms. The Fourier transforms of the sampling functions are toothcombs of equidistant $\delta-$functions; again, for the x-direction, they are
\begin{equation}
\tilde{\Sh}(u;\pixel)=\sum_{n=-\infty}^{\infty} \delta\left(u-\frac{n}{\pixel}\right)\;.
\end{equation}
 
In this work, we will assume that the brightness of the image drops to zero at the edge of the field of view, i.e., we will consider the limit $\Delta\theta\rightarrow\infty$. In this case, $\tilde{W}_{\rm fov}(u)=\tilde{W}_{\rm fov}(v)=\delta(u)\delta(v)$ and the Fourier transform of equation~(\ref{eq:discr}) simply becomes
\begin{equation}
V_{\rm m}(u,v)=\sum_{n=-\infty}^{\infty}\sum_{m=-\infty}^{\infty} V\left(u-\frac{n}{\pixel},v-\frac{m}{\pixel}\right)
\label{eq:Vm}
\end{equation}
This equation describes the effect of discretization on the calculated visibilities of the model image and demonstrates the aliasing introduced by the process. 

If the true image is Nyquist-sampled, i.e., if it has negligible power at baselines larger than $(2\pixel)^{-1}$, 
then there are no aliasing effects; only the $m=n=0$ terms contribute to the sum in equation~(\ref{eq:Vm}). However, as we discussed in detail in the introduction and will demonstrate in the next section, the images of black-hole shadows observed with the EHT are not Nyquist-sampled when the chosen pixel spacing is only marginally smaller than the nominal resolution of the EHT.

\begin{figure}[t]
  \centerline{  \includegraphics[width=0.49\textwidth]{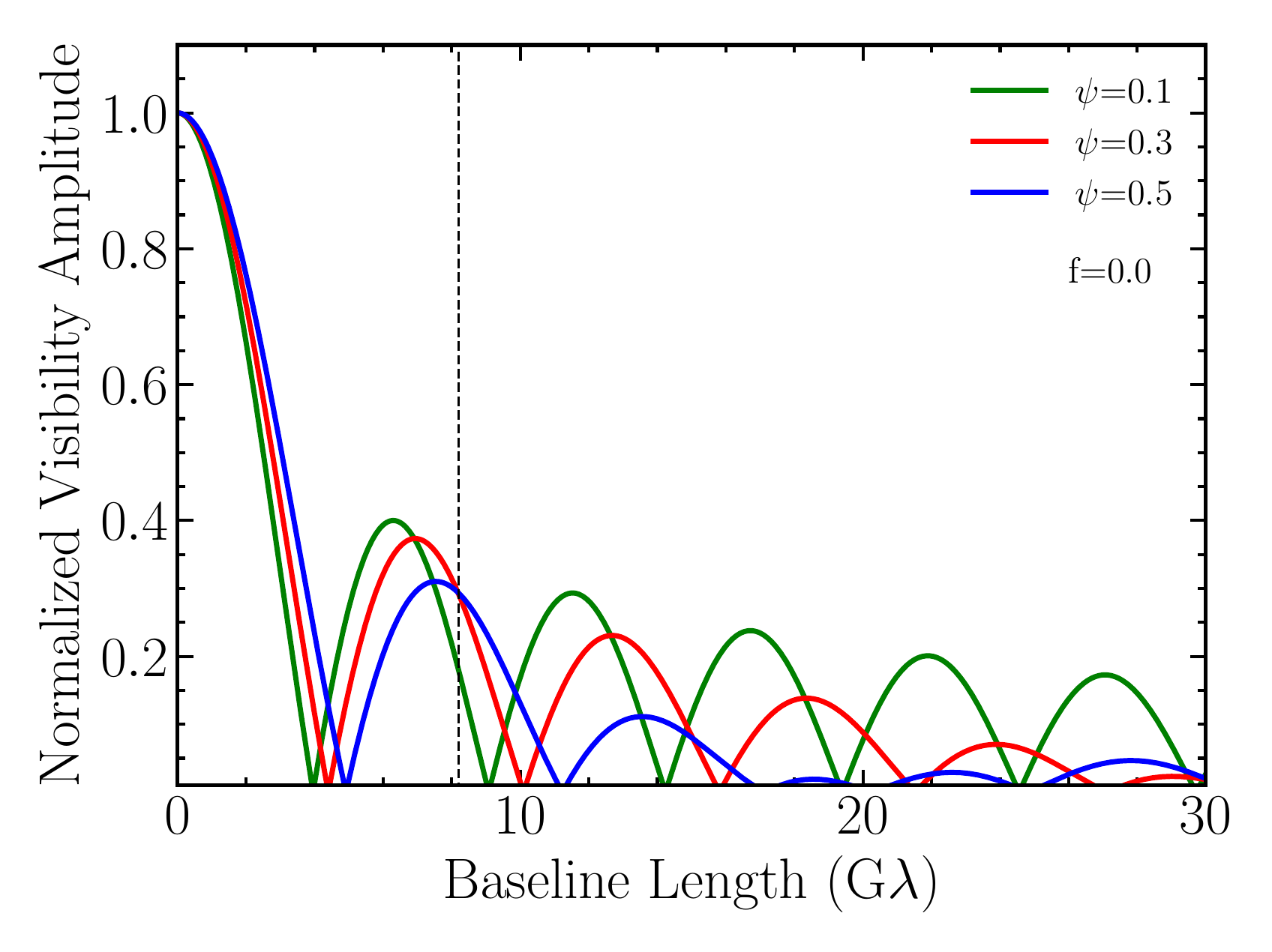}}
    \caption{The dependence of normalized visibility amplitude on baseline length for symmetric rings with different fractional widths $\psi$. The outer radius of the ring is set to $R_p=21~\mu$as, to resemble the 2017 EHT image of M87. The vertical dashed line marks the largest EHT baseline for the 2017 observations; the sharp features of the image generate substantial Fourier power at larger baseline lengths.
    \label{fig:ring_width}}
\end{figure}

\section{Optimal Pixel Spacing}

\subsection{Analytic Model Images}

As a first analytical estimate of the effects of aliasing on the visibilities of discretized model images of black-hole shadows, we will use the geometric model of a crescent image devised by~\citet{Kamruddin2013}. In this model, a crescent is generated by subtracting two uniform disks, with an asymmetry introduced by displacing the centers of the two disks with respect to each other. The model was generalized in~\citet{Benkevitch2016} with the introduction of a gradient in the brightness across the crescent, while further complexities were later added by allowing for a finite brightness depression in the center of the image and the addition of various displaced Gaussian components~\citep{PaperVI}. 

The first EHT data on M87 suggest a rather simple, ring-like structure with mild asymmetry~(see~\citealt{PaperVI}). For this reason and to reduce the complexity of our analytic estimates, we will only consider the case of a symmetric ring, generated by the subtraction of two uniform disks. The complex visibility of such an image is given by~\citep{Kamruddin2013}
\begin{equation}
V(u,v)=2\pi I_0 \left[\frac{R_p J_1(2\pi b R_p)}{2\pi b R_p}-\frac{R_n J_1(2\pi b R_n)}{2\pi b R_n}\right]\;.
\label{eq:Vring}
\end{equation} 
Here $R_p$ and $R_n$ are the outer and inner disk angular radii, $J_1(x)$ is the Bessel function of the first kind, $I_0$ is the uniform brightness of the ring, and 
\begin{equation}
b\equiv \sqrt{u^2+v^2}\;,
\end{equation}
is the baseline length.
Following~\citet{Kamruddin2013}, we will also express the fractional width of the ring in terms of the quantity
\begin{equation}
\psi\equiv \frac{R_p-R_n}{R_p}\;.
\end{equation}
Finally, we will normalize the complex visibilities such that they are equal to unity at zero baseline length, i.e., set 
\begin{equation}
I_0=\frac{1}{\pi R_p^2-(1-f)\pi R_n^2}\;.
\end{equation}

Figure~\ref{fig:ring_width} shows the visibility amplitude as a function of baseline length for rings of various fractional widths. As it is well known, the visibility amplitude of a uniform ring of infinitesimal width ($\psi=0$) shows a deep minimum at a baseline length of $\simeq 0.383/R_p$ (see~\citealt{TMS2017}). For the EHT observations of M87, this deep minimum occurs at a baseline length of $b_0\simeq 3.75 G\lambda$, which corresponds to a ring radius of~\citep{PaperVI}
\begin{equation}
R_p\simeq 21 \left(\frac{3.75\,{\rm G}\lambda}{b_0}\right)~\mu{\rm as}\;.
\end{equation} 
The maximum baseline of the 2017 EHT observations of M87 was $\sim 8.2$~G$\lambda$~\citep{PaperIII}, which is comparable to the location of the second visibility minimum. As Figure~\ref{fig:ring_width} shows, the expected visibility spectrum of the underlying image has substantial power at baselines larger than the location of the second minimum, i.e., than the larger baseline of the EHT array. 

The fractional width of the emission ring in the M87 images was not well constrained from imaging observations. However, visibility-domain modeling of the EHT data, especially on the first day of observations, suggests a rather small fractional width of $\psi\simeq 0.1$ (see Fig.~16 of~\citealt{PaperVI}). The images reconstructed from the EHT data with regularized maximum likelihood methods used $2~\mu$as pixels~\citep{PaperIV}, while the images of GRMHD simulations used $1~\mu$as pixels~\citep{PaperV}; these correspond to fractional pixel spacing of $\delta\theta/R_p\sim 0.1$ and $0.05$, respectively, that is comparable to the probable fractional width of the ring. Finally the fractional errors in the measurements of the ALMA baselines in the array were often better than 1\%~\citep{PaperIII}. These three sets of values set the benchmarks for our investigation below.

Using this analytic model, we calculate the fractional error between the true visibility $V(u,v)$ given by equation~(\ref{eq:Vring}) and the visibility of a digitized image $V_{\rm m}(u,v)$ given by equation~(\ref{eq:Vm}) as
\begin{equation}
\epsilon= \left\vert\frac{V_{\rm m}(u,v)}{V(u,v)}-1\right\vert\;.
\label{eq:biasan}
\end{equation}
For a ring of fixed size $R_p$, this error depends only on the fractional width of the ring $\psi$ and the ratio $\delta\theta/R_p$ between the pixel spacing $\delta\theta$ and the characteristic size of the ring $R_p$. 

\begin{figure}[t]
  \centerline{  \includegraphics[width=0.49\textwidth]{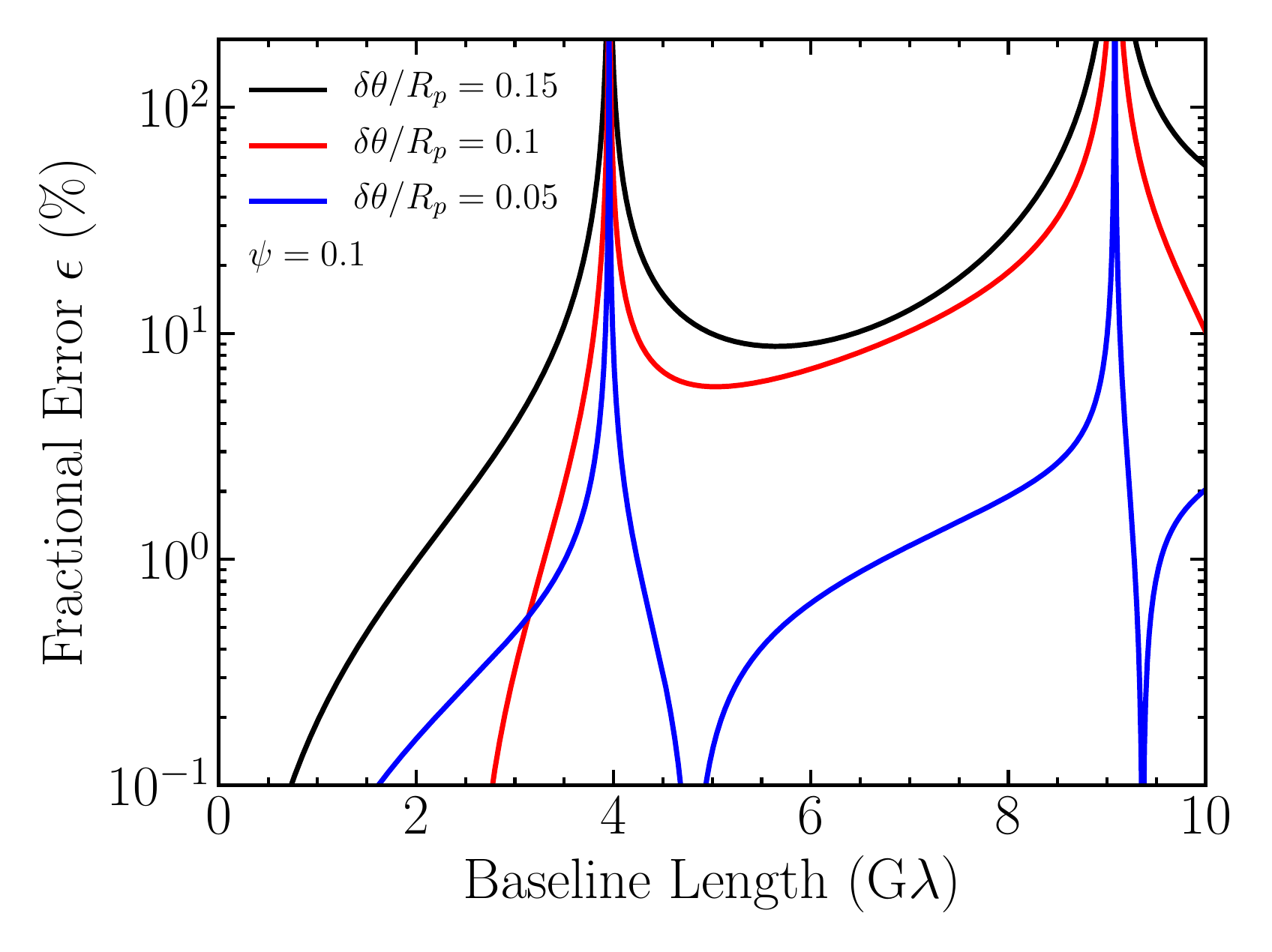}}
    \centerline{  \includegraphics[width=0.49\textwidth]{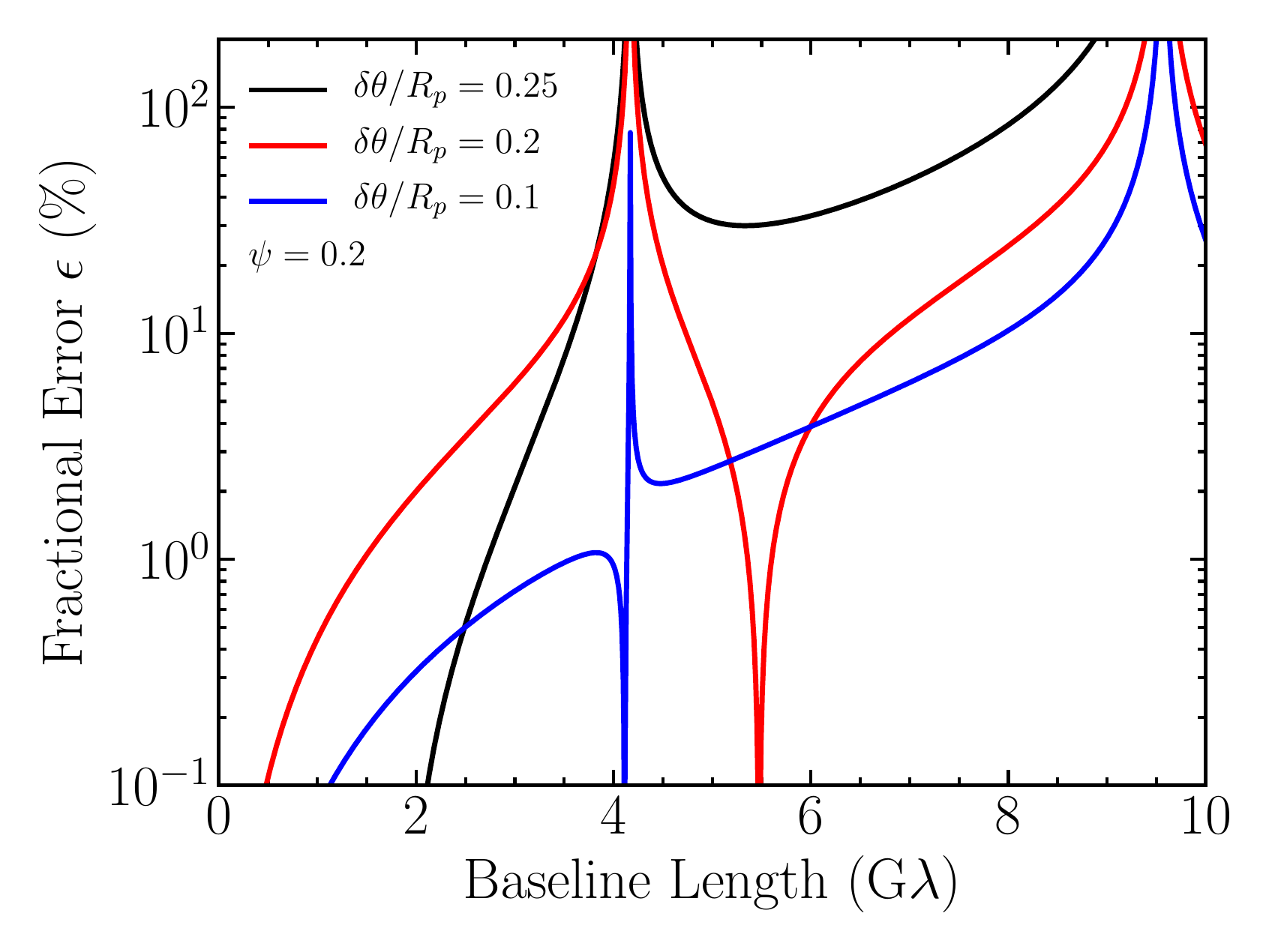}}
    \caption{ The fractional error $\epsilon$ between the visibility amplitudes of continuous crescent images and discretized ones. In both panels, the characteristic scale of the crescent is $R_p=21~\mu$as to resemble the 2017 EHT image of M87. The fractional width is set to {\em (Upper panel)\/} $\psi=0.1$ and {\em (Lower panel)\/} $\psi=0.2$. The results for three different pixel spacings (in units of $R_p$) are shown. As expected, the error is maximum near the visibility minima but can be as large as $\gtrsim 10$\% at intermediate baselines. The error, even at intermediate baseline lengths, is suppressed only when the pixel width resolves the smallest scales in the image that carry substantial power.
\label{fig:ring_bias_analytic}}
\end{figure}

\begin{figure*}[t]
  \centerline{  \includegraphics[width=0.49\textwidth]{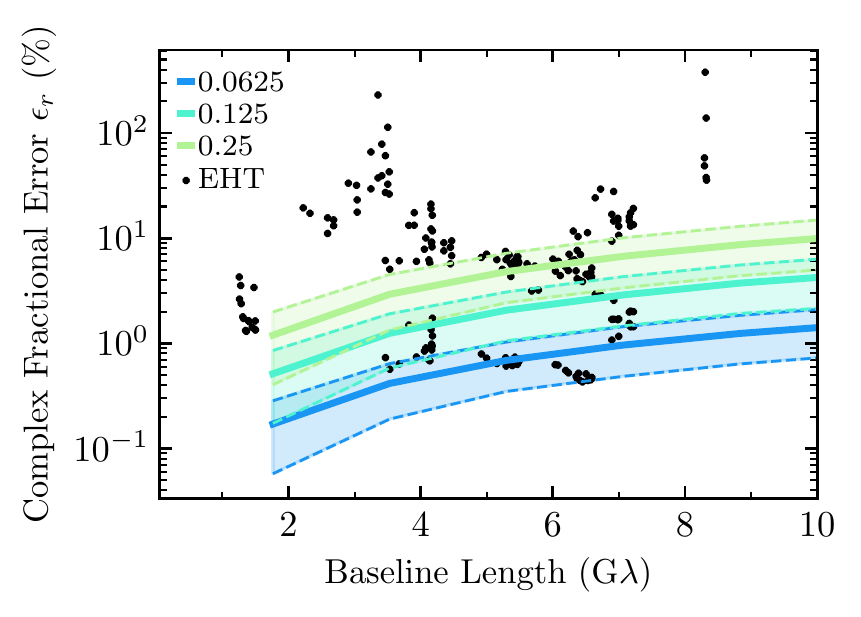}
    \includegraphics[width=0.49\textwidth]{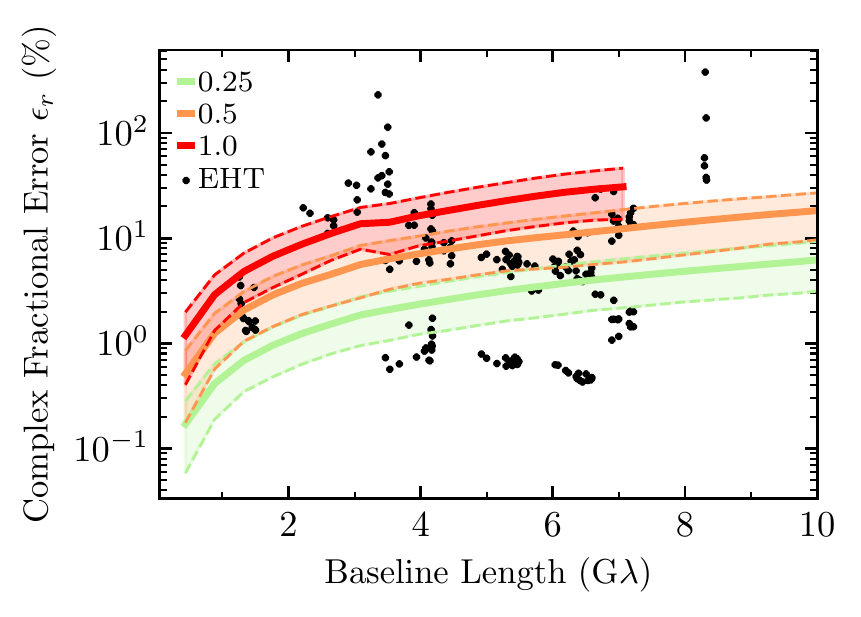}}
    \caption{The fractional error $\epsilon_r$ of the visibility amplitudes of model M87 images generated from a GRMHD simulation with different pixel spacings and different fields-of-view. The two panels correspond to fields-of-view of {\em (left)\/} 32  and {\em (right)\/} 128$GM c^{-2}$; for comparison, the diameter of the black-hole shadow is $\sim 10 GM c^{-2}$. For each panel, the visibilities of images with different pixel spacings (shown in the legend in units of $GM c^{-2}$)) were compared to those of the highest resolution images with $N=1024$ pixels along each direction. The shaded areas show the 68-percentile range of errors along different orientations in the $u-v$ plane, as a function of baseline length. The black points show the uncertainties of the 2017 EHT measurements. For the discretization to introduces errors that are less than a few percent requires a pixel spacing that is smaller than $\simeq 0.125 GM c^{-2}$. 
\label{fig:GRMHD_bias}}
\end{figure*}

Figure~\ref{fig:ring_bias_analytic} shows the fractional error between the visibility amplitudes of the continuous model image and the discretized one for crescents of two different fractional widths ($\psi=0.1$ and $\psi=0.2$) and for different pixel spacings. As expected, in all cases, the fractional error is larger at the location of the visibility minima, where the model amplitude drops to negligible values. Away from the locations of the visibility minima, the error is at the $\sim 10$\% level, which is larger than the formal measurement error in the EHT data. It is important to emphasize here that structures at scales as small as 2~$\mu$as, which nominally require baseline lengths as large as $\simeq 100$~G$\lambda$ to be resolved, introduce aliasing at the 10\% level at baseline lengths that are 20 times smaller and, therefore, at image scales that are 10 times larger than the scale of these structures. The error becomes suppressed only when the pixel spacing is small enough to resolve the smallest scale in the image that carries substantial power, which in this case is the width of the crescent.

\subsection{Numerical Model Images}

In model images generated by GRMHD simulations, the minimal scale that will need to be resolved is determined by the sharpness of the flux suppression at the black-hole shadow and the power of small-scale filamentary structures in the plasma. In order to assess this, we explored and contrasted the visibility amplitudes of a number of model images with different fields-of-view and pixel spacings. 

In principle, the characteristic scales in the GRMHD images that carry significant power depend also on the resolution of the simulations themselves (see, e.g.,~\citealt{Porth2019}). However, because of the stochastic nature of turbulence, running the GRMHD simulations at different resolutions does not simply resolve filamentary structures better but rather generates a completely different set of structures. As a result, assessing the dependence of the bias on the resolution of the GRMHD simulations can only be done in a statistical way by using a large ensemble of images from these simulations. This is beyond the scope of the current study, which only focuses on the bias introduced by the discretization of the images for a given GRMHD resolution.

We used the GRMHD simulations+GR radiative transfer models of \citet{Chan2015}, choosing parameters that are relevant to the M87 images observed by the EHT. In particular, we used the GRMHD simulation with a black-hole spin $a=0.9$ and a Standard And Normal Evolution (SANE) magnetic field structure~\citep{Narayan2012}  to describe the accretion flow (the \texttt{a9SANE} model of~\citealt{Chan2015}). For the radiative transfer, we set the mass of the black hole to $M = 6.5\times10^9~M_\odot$, the inclination of the observer to $i =17^{\circ}$, and the electron temperature to follow the $\beta$-law with $ R_{\mathrm{high}} = 20$ (see~\citealt{PaperV}). We performed the radiative transfer calculations at a wavelength of 1.3mm using the GPU-accelerated GRay algorithm~\citep{Chan2013}. In order to calculate angular sizes in the sky, we used a distance of 16.8 Mpc.

We explored three fields of view ($128GMc^{-2}$, $64GMc^{-2}$, and $32GMc^{-2}$) and, for each field of view, four resolutions corresponding to $128,\,256,\,512,\,\mathrm{and}\,1024$ pixels, respectively. The combination of the smallest field of view and the largest number of pixels gave rise to the smallest pixel spacing in our set ($32 GM c^{-2}/1024=0.03125~GM c^{-2}$), while the combination of the largest field of view and the smallest number of pixels gave rise to the largest pixel spacing in our set ($128 GM c^{-2}/128=GM c^{-2}$). We compared the discretized complex visibilites $V_{ij}$ of images with different resolutions or fields-of-view to those of the images with the best resolution, $V_{ij,0}$. We then calculated the error due to discretization as
\begin{equation}
\epsilon_r\equiv \left\{ \frac{[V_{ij}-V_{ij,0}][V^*_{ij}-V^*_{ij,0}]}{\vert V_{ij,0}\vert^2}\right\}^{1/2}\;.
\end{equation}
Here, a star superscript denotes the complex conjugate of the complex visibility and we wrote this expression explicitly in order to emphasize the fact that we are calculating complex differences and not differences of visibility amplitudes.

\begin{figure*}[t]
  \centerline{\includegraphics[width=0.9\textwidth]{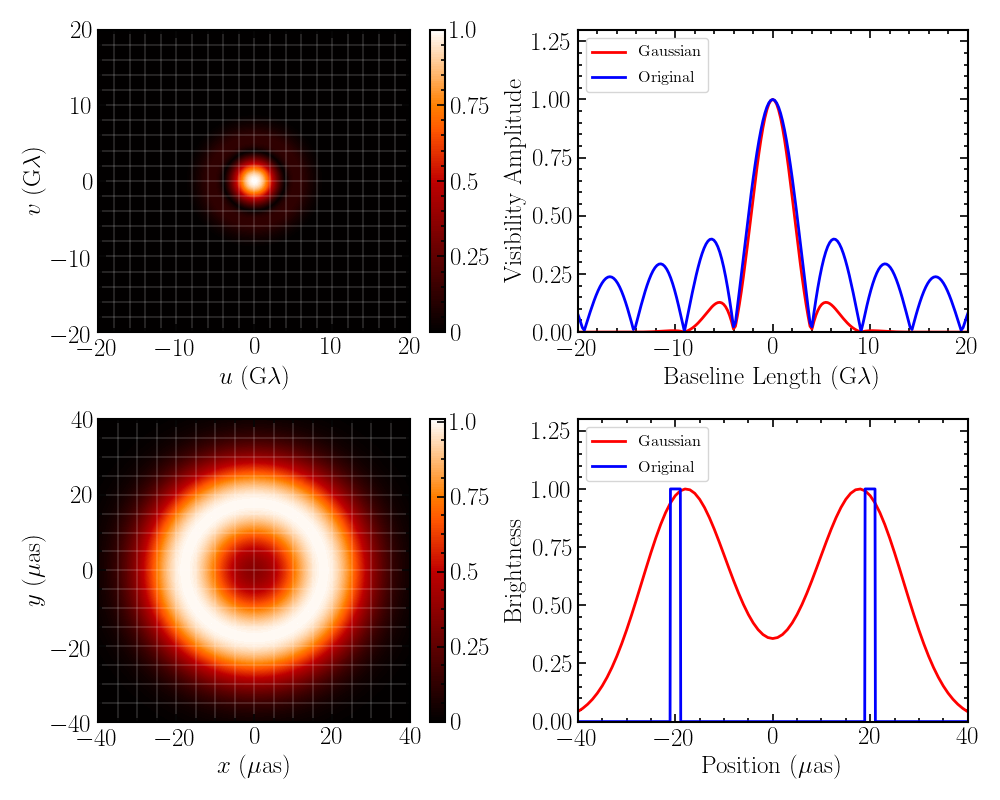}}
    \caption{{\em (Left Panels)\/} The $u-v$ map and the corresponding image of a ring shape that has been smoothed by a Gaussian with a 20$ \mu$as FWHM, as was done in the EHT images of M87. The ring model has an outer radius of 21~$\mu$as and a fractional width of $\psi=0.1$ to mimic the M87 image. {\em (Right Panels)\/} Horizontal cross sections of the $u-v$ map and of the image of the ring. The blue curves show the original (unfiltered) visibilities and image and the red curves show the filtered ones. The Gaussian filter suppresses the power of structures even at the $5-10$~G$\lambda$ baseline lengths, where high quality data exist, and biases the size of the bright ring towards small values. 
\label{fig:gauss_filter}}
\end{figure*}

Figure~\ref{fig:GRMHD_bias} shows the error introduced by different pixel spacings in simulations with different fields-of-view, as a function of baseline length. The shaded areas in each figure outline the 68-percentile range of error calculated along different orientations in the $u-v$ plane at a given baseline length. The error is also compared to the uncertainties in the measurements of the M87 visibilities with the 2017 EHT array~\citep{PaperIII}. Because the model images are very compact, the error depends very weakly on the field-of-view, which is why we only show the result for the smallest and the largest fields-of-view that we considered. However, as expected from the analytic considerations discussed above, the error at all baselines increases rapidly with pixel spacing. For the set of simulated images we considered in this study, the bias is smaller than a few percent for baselines $\lesssim 8$G$\lambda$ and hence comparable to the smallest uncertainties in the EHT data only when the pixel spacing is $\lesssim 0.125 GM c^{-2}$. For both of the primary EHT targets, this corresponds to an angular spacing of $\lesssim 0.5~\mu$as.

\bigskip

\bigskip

  \section{Best Filtering Practices}
 
 As discussed in the introduction, images generated from interferometric data using different techniques are commonly smoothed at a scale comparable to the nominal resolution of the instrument in order to suppress any small-scale structures that the array cannot resolve. A similar filtering is performed when comparing the high-resolution model images from GRMHD simulations, which are not constrained by the resolution of the array, to those from imaging algorithms.

There is a large literature of filter design for digital signal processing (see, e.g., \citealt{Harris1978} for an early review). The optimal filter depends on the requirements of the particular application in which the filter is used. The filtering requirement for the case of images obtained with an interferometric array is qualitatively different than traditional filtering in image processing, which often aims to remove high-frequency noise from images.  Indeed, even at large baselines ($\sim 7$~G$\lambda$; see Fig.~\ref{fig:GRMHD_bias}), the EHT data are not typically dominated by noise. For this reason, filtering with Wiener-type filters is not directly applicable in the case we consider here. The only exception is with data from Sgr~A*, the black-hole in the center of the Milky way, for which the largest baselines are expected to be dominated by refractive noise~\citep{Psaltis2018, Johnson2018}.

Traditionally, Gaussian filters have been used in interferometric imaging because they do not introduce any artifacts (often called the ``bokeh'' effect) in images. The Fourier transform of a Gaussian is also a Gaussian and, therefore,  convolving it with an underlying image causes simple blurring. However, applying a Gaussian filter suppresses the Fourier components of the image at all frequencies, even at those for which accurate data exist. In other words, interferometric images that have been blurred with a Gaussian filter show substantially less structure than what is encoded in the visibility data that they are meant to describe. 

\subsection{Analytic Model Images}

As a concrete example, we explore geometric models that are similar to the first EHT images of M87 and have been smoothed with a Gaussian filter of FWHM equal to $20 \mu$as~\citep{PaperI}. Figure~\ref{fig:gauss_filter} shows the $u-v$ map and the corresponding image of a geometric ring with outer radius $R_p=21~\mu$as and fractional width $\psi=0.1$ (see \S2) that has been smoothed with such a Gaussian filter. The filter introduces a factor of $\sim 5-10$ suppression of the visibility amplitudes even at the $\sim 5-10$~G$\lambda$ baselines for which high quality data exist. Moreover, the filter biases the size of the bright ring towards small values; this bias has been identified as one of the reason for the apparent discrepancy between the ring sizes inferred for M87 using image-domain and visibility-domain methods~\citep{PaperIV, PaperVI}.

 \begin{figure}[t]
  \centerline{  \includegraphics[width=0.49\textwidth]{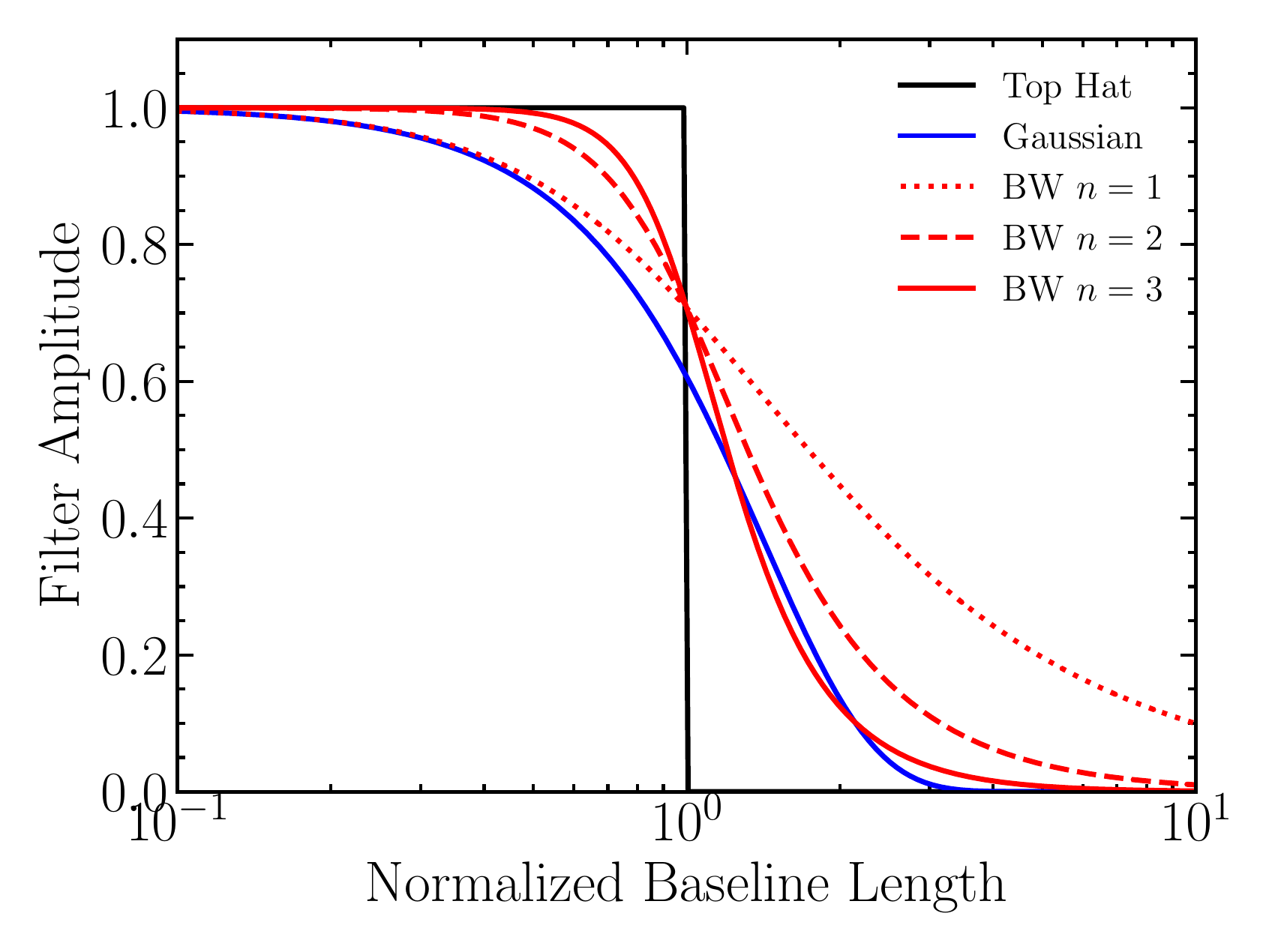}}
    \caption{The response of {\em (black)} a top-hat filter, {\em (blue)} a Gaussian filter, and {\em (red)} the three lowest-order Butterworth filters. The half-width of the top-hat, the dispersion of the Gaussian, and the scale $r$ of the Butterworth filters were all set to unity. Among them, the Butterworth filter is optimal because it combines flat response over a large range of short baseline lengths while introducing a shallower decline to zero at large baselines to reduce ringing. 
\label{fig:filters}}
\end{figure}

\begin{figure}[t]
  \centerline{  \includegraphics[width=0.49\textwidth]{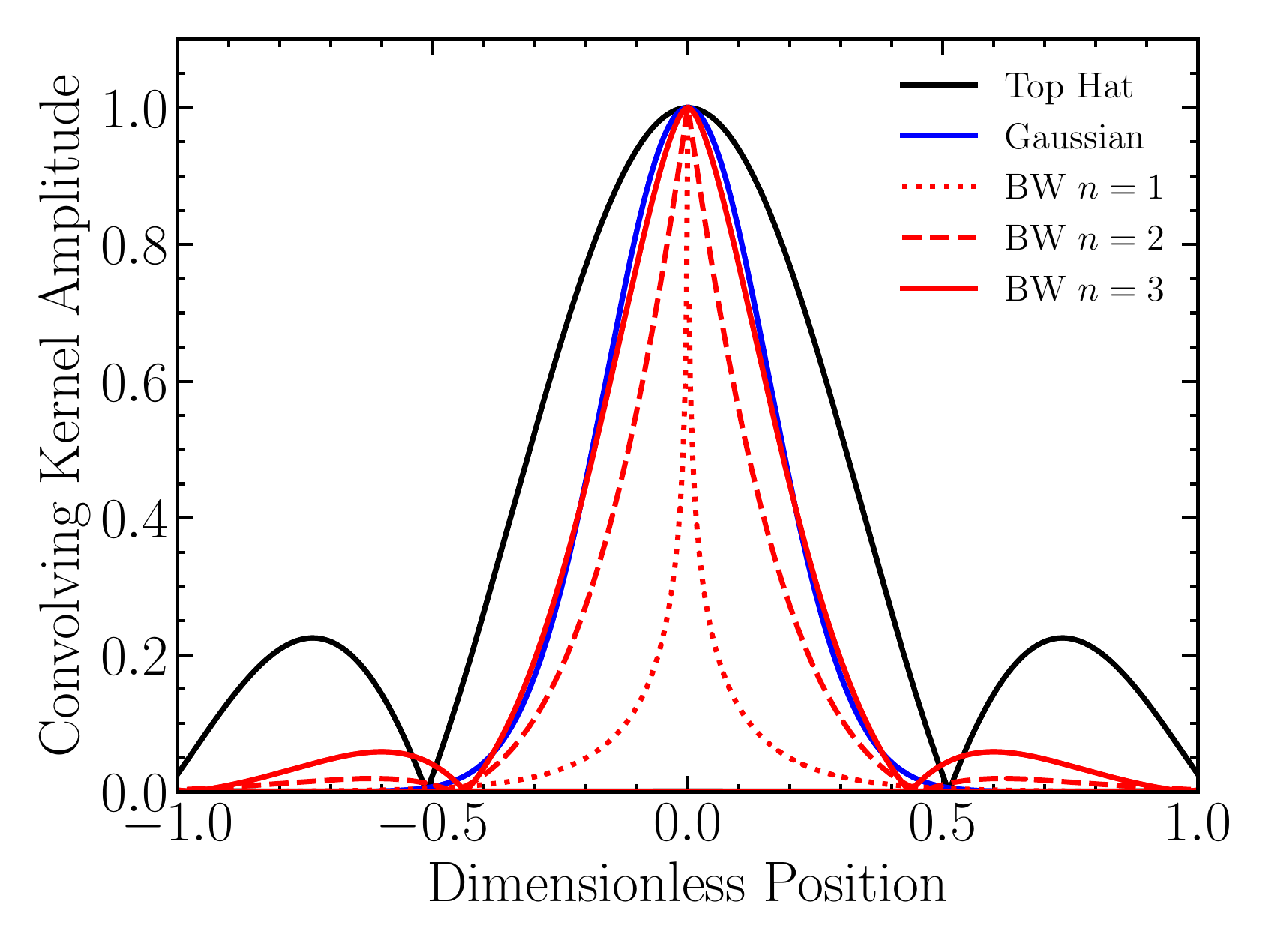}}
    \caption{The image-domain kernels of the five filters shown in Figure~\ref{fig:filters}. Applying each of the filters to an image is equivalent to convolving the image with the kernels shown here. Increasing the power-law index of the Butterworth filter beyond $n=2$ only marginally affects the width of the convolving kernel but increases the amplitudes of the sidelobes. The $n=2$ Butterworth filter is optimal because of its flat low-frequency response and the minimal amplitude of the sidelobes.
\label{fig:filters_conv}}
\end{figure}

The response of an optimal filter for our purposes will need to satisfy three conditions: {\em (i)\/} be near unity at all baselines for which data exist, {\em (ii)\/} drop to zero quickly above the maximum baseline of the array, but {\em (iii)\/} have a shallow enough gradient in order not to introduce artifacts (ringing) in the presence of sharp structures like black-hole shadows. This last condition also ensures that flux does not artificially "leak in" the center of the image and contaminates potential signatures of the black-hole shadow. In the theory of signal processing, the Butterworth filter~\citep{Butterworth1930}
\begin{equation}
F_{\rm BW} (b)= \left[1+\left(\frac{b}{r}\right)^{2n}\right]^{-1/2}
\end{equation}
optimizes a flat low-frequency response and a smooth decline to zero at large Fourier frequencies. The Butterworth filter has two parameters: the scale $r$ and the power-law index $n$. 

\begin{figure*}[t]
  \centerline{\includegraphics[width=0.9\textwidth]{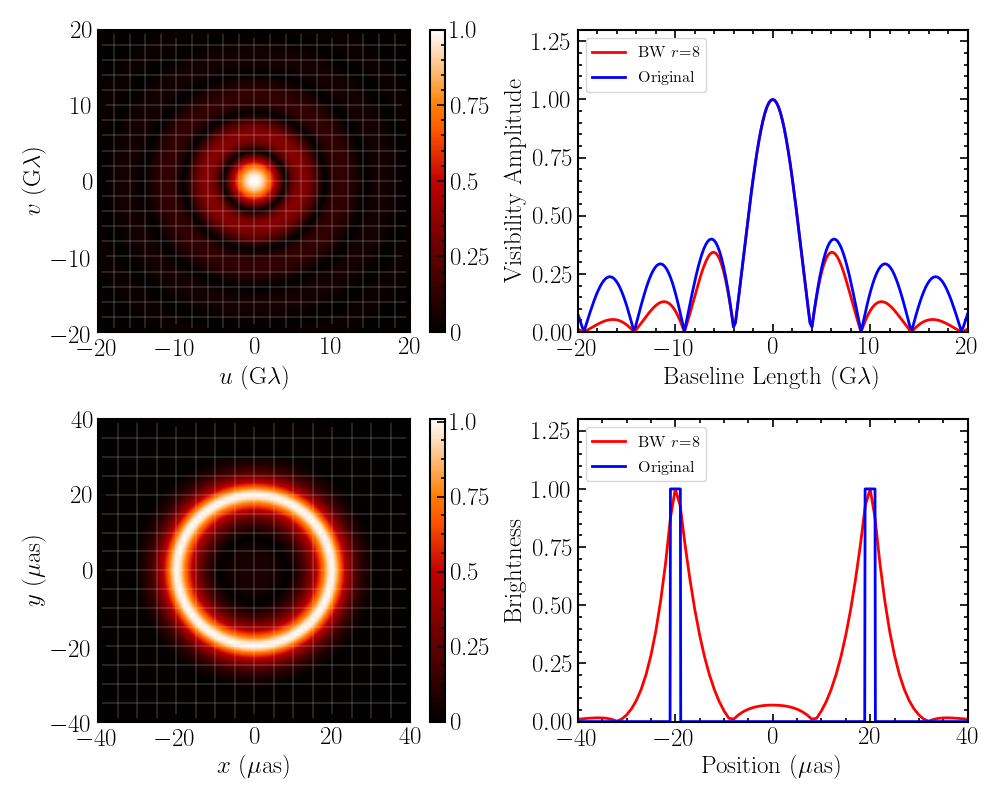}}
    \caption{Similar to Figure~\ref{fig:gauss_filter} but for a Butterworth filter with $n=2$ and a scale $r=8$~G$\lambda$ that is equal to the largest baseline of the EHT coverage. This filter effectively suppresses power only at baselines $\gtrsim 10$~G$\lambda$, where no data exist, leads to minimal bias on the size of the ring, and introduces artifacts only at the few percent level. This filter is appropriate for application to reconstructed images of black-hole shadows with the EHT and image-domain feature extraction algorithms.
\label{fig:bw_8}}
\end{figure*}

Figure~\ref{fig:filters} compares the Butterworth filter for different values of the power-law index $n$ to the Gaussian and to the top-hat filters, in the visibility (Fourier) domain. For cross-comparison, the characteristic scale of each filter, i.e., the half-width of the top-hat, the dispersion of the Gaussian, and the scale $r$ of the Butterworth filter are set to one. Figure~\ref{fig:filters_conv} compares the filters in the image domain, i.e., the kernels with which the images will be convolved. As discussed above, the Gaussian filter has a very shallow dependence on baseline length but the corresponding convolving kernel reduces smoothly to zero, with no ringing. In the opposite extreme, the top-hat filter has a very abrupt transition from unity response to zero response but the corresponding convolving kernel shows large sidelobes and introduces substantial ringing. Changing the power-law index $n$ of the Butterworth filter allows us to explore filters with intermediate properties and optimize their parameters. 

The $n=1$ Butterworth filter corresponds to a very narrow, centrally peaked convolving kernel with no sidelobes and, hence, no ringing. However, this occurs because the $n=1$ filter has an even shallower decline than the Gaussian filter. As a result it does not satisfy our first two requirements. As the power-law index of the Butterworth filter is increased, its shape becomes closer to the top-hat function and, consequently, the amplitudes of the sidelobes also increase. 

The $n=2$ Butterworth filter offers the best compromise between a top-hat-like filter in Fourier space and negligible ringing in image space. The amplitude of the filter at a baseline equal to $r$ is equal to $0.71$ and it drops to $0.11$ at $3r$. In other words, if we choose the scale of the filter to be equal to the largest baseline in the array, then the filter will preserve practically all the information in the image at scales up to those probed by the array while suppressing by a factor of 10 any power at scales three times smaller. At the same time, the maximum kernel amplitude of this filter at the location of the sidelobe is only $\sim 2$\%. Therefore, even for an infinitely sharp flux depression at the location of the black-hole shadow, the $n=2$ Butterworth filter will only introduce artifacts (ringing) at the few percent level.

Figure~\ref{fig:bw_8} shows the $u-v$ map and the corresponding image of the same geometric ring as that in Figure~\ref{fig:gauss_filter} but when the $n=2$ Butterworth filter is applied to the image with a scale $r=8$~G$\lambda$ equal to the maximum baseline of the EHT array. As expected from the above discussion, the filter effectively suppressed most of the power at baselines larger than $r$ while retaining the shape of the visibility spectrum at baselines $\lesssim r$, where data exist. The sidebands of the convolving kernel fill in the central brightness depression at the few percent level, which is substantially smaller than the leakage from the Gaussian filter (see Fig.~\ref{fig:gauss_filter}). Moreover, this Butterworth filter introduces negligible bias to the location of the brightness maxima and, therefore, to the inferred size of the ring. For this reason, this filter is appropriate for application to reconstructed images of black-hole shadows and of image-domain feature extraction algorithms (see, e.g.,~\citealt{PaperVI}).

\begin{figure}[t]
  \centerline{  \includegraphics[width=0.49\textwidth]{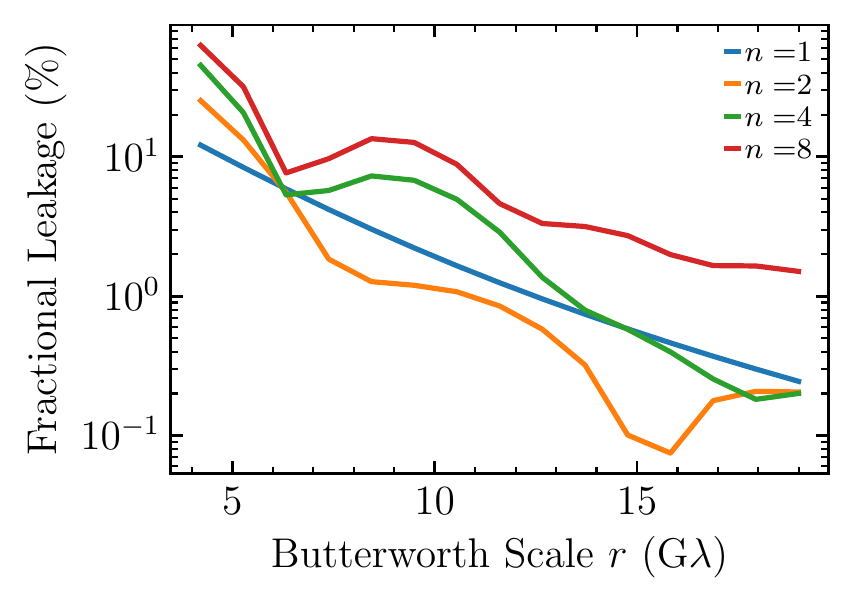}}
    \caption{The fractional leakage of flux in the center of the black-hole shadow for the image shown in Figure~\ref{fig:grmhd_filter} caused by the sidelobes of the Butterworth filter as a function of the filter scale $r$ for different values of the power-law index $n$. Even for a feature as sharp as the black-hole shadow, this artifact of filtering is at the $< 1$\% level when we choose the $n=2$ Butterworth filter with a scale $r=15$~G$\lambda$.
\label{fig:central_bright}}
\end{figure}

\subsection{Numerical Model Images}

Even though the parameters of the Butterworth filter discussed above introduce artifacts at the $\lesssim 5$\% level only, there are other use cases for which one is interested in filtered images with substantially smaller artifacts. An example of such a case is the application of Principal Component Analysis (PCA) to a large suite of GRMHD model images to identify a minimal set of eingenimages that can then be used in modeling the observations in the visibility domain~\citep{Medeiros2018a}. The number of eigenimages in this minimal set can be reduced by applying PCA only to images for which the small-scale details have been filtered away. However, filtering the images should not introduce artifacts that could alter their Fourier transforms at a level larger than the uncertainties in the interferometric data.

In order to explore in more detail such an application of Butterworth filters to GRMHD model images, we used a high-resolution image of the simulation we discussed in \S3.2.  We convolved the model image with Butterworth filters of different scales $r$ and power-law indices $n$. We then measured the mean brightness within a square of size $2.5 GM c^{-2}$ that surrounds the center of the shadow and defined a fractional leakage by dividing this number by the maximum brightness of the filtered image. 

Figure~\ref{fig:central_bright} shows the fractional leakage as a function of the filter parameters. As expected from the analytical discussion above, increasing the order of the Butterworth filter increases the amplitude of the sidebands of the convolving kernel and, therefore, the amount of leakage. At the same time, increasing the scale $r$ causes the convolving kernel to become narrower and leads to an overall decrease of the fractional leakage. 

The local maximum at $r\simeq 10$~G$\lambda$ for $n=2$ is a consequence of the fact that the black-hole image has a ring-like structure with a radius of $\simeq 20~\mu$as. As can be seen from Fig.~\ref{fig:filters_conv}, the sideband of the Butterworth filter occurs at a relative position of $\sim 0.6/r$. As a result, for $r=10$~G$\lambda$, the sideband will appear $\simeq 20~\mu$as away from each maximum in the image causing the sidebands of all maxima around the perimeter of the photon ring to add up at the center of the ring.

For the typical black-hole image used here, the $n=2$ Butterworth filter generates the least amount of fractional leakage because it combines a narrow convolving kernel (in contrast to the $n=1$ filter) and small sidebands (in contrast to the $n>2$ filters). Requiring the fractional leakage to be $\lesssim 1$\%, for a black hole shadow with a size comparable to that of M87, places a lower limit on the filter scale of $r\gtrsim 13$~G$\lambda$.

\begin{figure*}[t]
  \centerline{\includegraphics[width=0.9\textwidth]{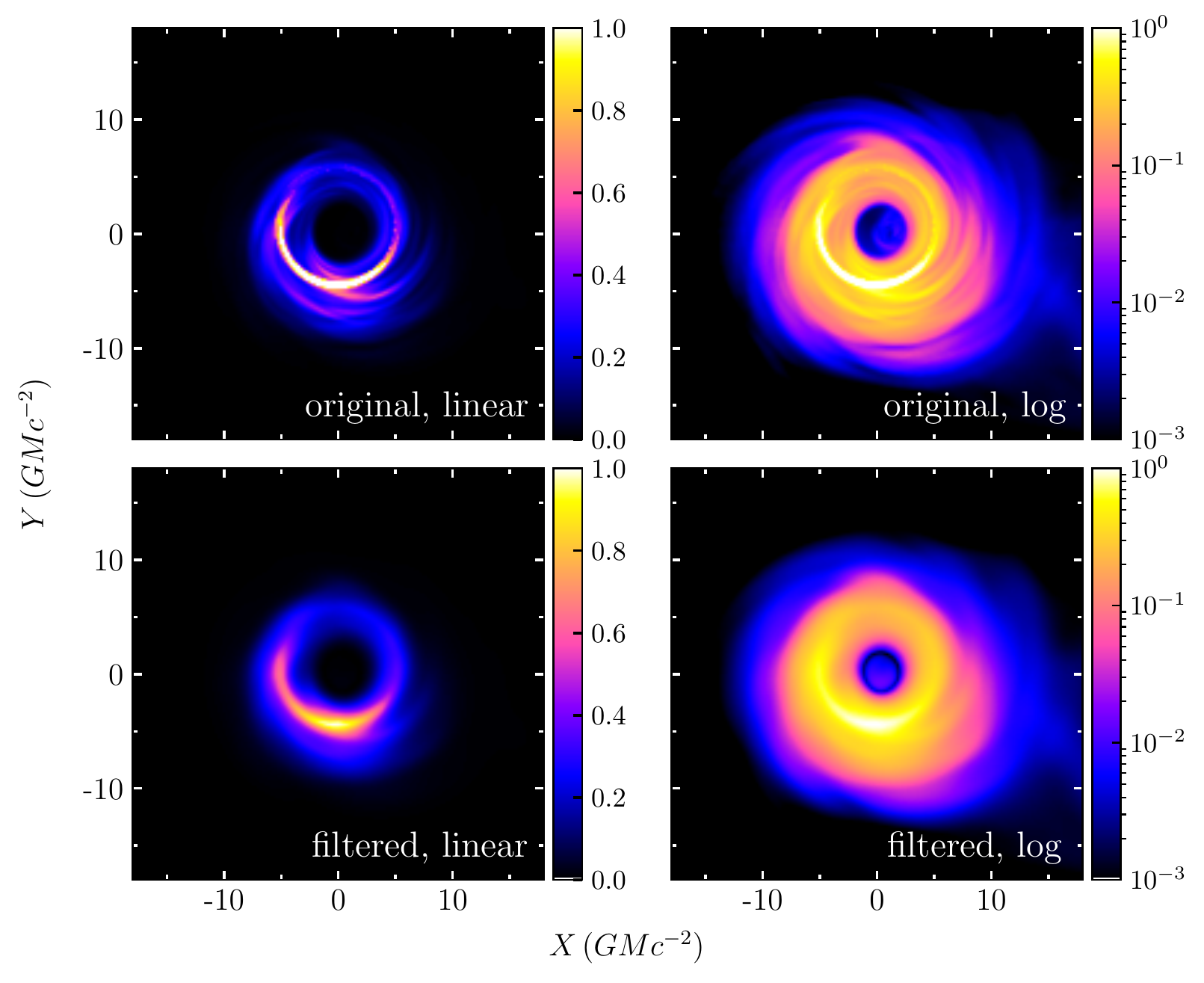}}
    \caption{{\em (Top Panels)\/} A model image from a GRMHD simulation shown in {\em (left)\/} linear and {\em (right)\/} logarithmic brightness scale; in both panels, the brightness is normalized to its maximum value throughout the image. {\em (Bottom Panels)\/} The same image after being filtered using a Butterworth filter with $r=15$~G$\lambda$ and $n=2$. Filtering preserves the large-scale structures seen in the image, even those with brightness equal to a few percent of the brightest pixels. The sidelobes of the filter introduce extraneous flux in the center of the black-hole shadow but only at the percent level. 
\label{fig:grmhd_filter}}
\end{figure*}

The top panels of Figure~\ref{fig:grmhd_filter} show the model GRMHD image we used in this exploration, in linear and in logarithmic brightness scale. The latter is important because it shows structures that have large scales and, therefore, need to be preserved, but have lower brightness. To represent the M87 image observed with the EHT, the spin axis of the black hole was set in the E-W orientation. The bottom panels of the same figure show the image after it has been convolved using a Butterworth filter with $r=15$~G$\lambda$ and $n=2$. 

\begin{figure*}[t]
  \centerline{\includegraphics[width=0.45\textwidth]{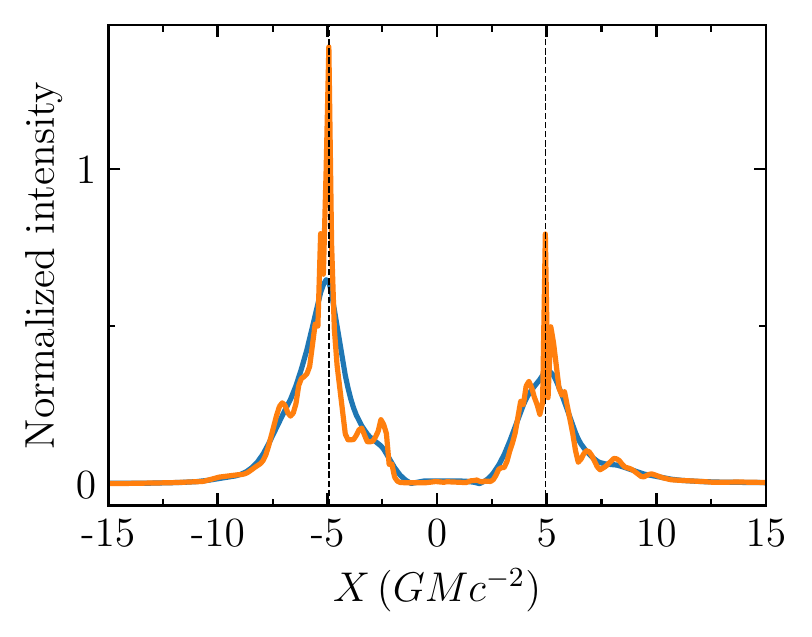}
  \includegraphics[width=0.45\textwidth]{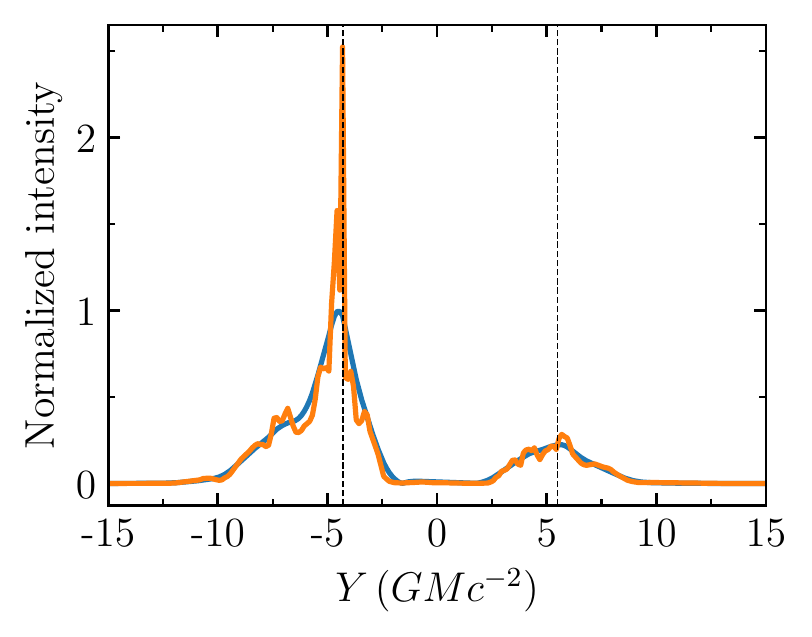}}
    \caption{{\em (Left)\/} Horizontal and {\em (Right)\/} vertical cross sections of the original (in orange) and filtered (in blue) images shown in Figure~\ref{fig:grmhd_filter}. The expected location of the bright ring surrounding the black-hole shadow is shown with the vertical dashed lines. The Butterworth filter removes effectively the sharp features of the image, without altering its overall shape or the size of the bright emission ring.
\label{fig:grmhd_filter_cross}}
\end{figure*}

As expected from the earlier discussion, the filter effectively removes the high frequencies while preserving the large-scale structures in the image. This is shown explicitly in Figure~\ref{fig:grmhd_filter_cross} that depicts a horizontal and a vertical cross section of the image through the center of the black-hole image, before and after the application of the filter. The location of the black-hole shadow for this black-hole spin is also shown, demonstrating that the narrow convolving kernel of the filter will introduce only negligible bias to the measurement of the shadow size of the black hole.

\section{Discussion}

At an abstract level, the technical task of the EHT is to measure a set of complex visibilities, or Fourier-domain components, of the images of nearby supermassive black holes in a way that is sufficient to resolve their shadows~\citep{PaperIII}.  The analysis of these interferometric data then aims to either construct an image from the visibilities~\citep{PaperIV} or construct a visibility model based on geometric expectations or GRHMD simulations that convincingly fit the data~\citep{PaperVI}.  

These two approaches, direct image reconstruction and visibility domain fitting, appear complementary,  at first glance. However, in both cases, the complex visibilities are measured in the Fourier domain, while the reconstructed images or simulations models are generated in the image domain.  Comparison of data in one domain with representations in the other can only be done correctly with rigorous respect for the mathematics of sampling and image formation.

The images generated with GRMHD simulations, in particular, instantly highlight the challenge of accurately mapping them onto the EHT observables.  For example, Figure~2 of~\citet{PaperV} vividly shows the effects of aliasing in the spatial domain, particularly at the sharp inner edge of the photon ring.   The jagged inner edge of the photon shadow clearly demonstrates that aliasing will certainly bias high spatial frequencies. However, the “hot” pixels are often spaced quite far apart, which means that erroneous low-frequency power has also been generated; the implied spatial scale is a large fraction of the shadow diameter,  which the EHT observations are sensitive to. Similar considerations apply to the seemingly more objective direct image reconstructions. 


Our first aim in this article was to quantify the minimum sampling frequency required such that aliasing does not degrade the low-frequency power above the errors in the corresponding EHT visibilities, or cause unacceptable biases in the parametric inferences derived when fitting simulations to the data. Using GRMHD simulations that are relevant to the parameters of M87, we estimated a maximum pixel spacing of $(1/8) GM c^{-2}$. For the mass and distance to both of the primary EHT targets, Sgr~A* and M87, this corresponds to an angular pixel spacing of $\lesssim 0.5\mu$as.

A direct consequence of the required small pixel spacing is the fact that the model and reconstructed images need to be convolved with appropriate filters in order to show only scales that are constrained by the actual interferometric data. It is important to emphasize here that filtering {\em cannot suppress strong  aliasing\/}; it just smears the damage around.   With a fine enough sampling scale, however, the extent of the alias is limited, understood, and can be eliminated with a proper cutoff filter. This is why our approach was to first control the aliasing with sufficiently fine sample and then eliminate it with a good cutoff filter.

We showed, using analytic models and GRMHD simulations, that an $n=2$ Butterworth filter with a scale equal to the largest baseline in the array does not suppress power at scales where data exist, diminishes quickly towards smaller scales, and introduces only marginal ringing because of the filter sidelobes. For applications in which the filtered images will be used to generate high-fidelity $u-v$ maps to compare directly GRMHD simulations to data, setting the scale of the Butterworth filter to $r\gtrsim 13$~G$\lambda$ ensures images with artifacts suppressed to the $\lesssim 1$\% level.

\acknowledgements

We thank  D. Ball, G. Bozzola, P. Christian, D. Heumann, L. Krapp, C. Raithel, A. Roshanineshat, K. Satapathy, and T. Trent for useful discussions. This work was supported in part by NSF PIRE grant 1743747. L. M. acknowledges support from an NSF Astronomy and Astrophysics Postdoctoral Fellowship under award no.\ AST-1903847.  All ray tracing and PCA calculations were performed with the El Gato GPU cluster at the University of Arizona that is funded by NSF award 1228509.

\bibliographystyle{apj}

\bibliography{pixels}

\end{document}